\documentclass[numberallpages]{llreport}

\usepackage[hide]{llmarkings}
\usepackage{tikz}  

\usepackage{adjustbox}
\usepackage{booktabs}
\usepackage{csvsimple}
\usepackage{tcolorbox}
\usepackage[T1]{fontenc}
\usepackage{placeins}
\usepackage{longtable}
\usepackage{xltabular}
\usepackage{makecell}
\usepackage{threeparttable}

\usepackage{tabularx}

\usepackage{caption}      
\usepackage{subcaption}

\usepackage{ifthen}

\newboolean{showdetails}
\setboolean{showdetails}{false} 

\usepackage{verbatim}     
\usepackage{listings}
\usepackage{listings-rust}
\usepackage{xcolor}

\definecolor{rustgrey}{RGB}{100,100,100}
\lstdefinestyle{ruststyle}{
	escapeinside={(*@}{@*)}, 
	language=Rust,
	numbers=left,
	stepnumber=1,
	numberstyle=\tiny\color{rustgrey},
	basicstyle=\ttfamily\small,
	breaklines=true,
	frame=single,
	commentstyle=\color{green},
	keywordstyle=\color{blue},
	stringstyle=\color{red},
}
\lstdefinestyle{cstyle}{
	escapeinside     = {(*@}{@*)}, 
	language=C,
	numbers=left,        
	stepnumber=1,        
	numberstyle=\tiny\color{gray}, 
	basicstyle=\ttfamily\small,
	keywordstyle=\color{blue},
	commentstyle=\color{green!60!black},
	frame=single,        
	breaklines=true      
}
\let\origthelstnumber\thelstnumber
\makeatletter
\newcommand*\stopnumber
{
	\lst@AddToHook{OnNewLine}
	{
		\let\thelstnumber\relax
		\advance\c@lstnumber-\@ne\relax
	}
}
\newcommand*\startnumber[1]
{
	\setcounter{lstnumber}{\numexpr#1-1\relax}
	\lst@AddToHook{OnNewLine}
	{
		\let\thelstnumber\origthelstnumber
		\refstepcounter{lstnumber}
	}
}
\makeatother

\usepackage{cleveref}
\usepackage{placeins} 

\usepackage{comment}
\usepackage{dirtree}
\usepackage{hyperref}

\usetikzlibrary{arrows,shapes,automata,backgrounds,petri,mindmap}

\usepackage{acronym}
\usepackage{xspace}
\acrodef{LLM}{Large Language Model}
\acrodef{I/O}{input/output}
\acrodef{JSON}{JavaScript Object Notation}
\acrodef{UB}{undefined behavior}

\newcommand{\wisc}{University of Wisconsin\xspace}
\newcommand{\wash}{University of Washington\xspace}
\newcommand{\galois}{Galois\xspace}
\newcommand{\aarno}{Aarno Labs\xspace}
\newcommand{\yale}{Yale University\xspace}
\newcommand{\intel}{Intel Corporation\xspace}

\newcommand{\performanceperformer}[1]{%
	\ifthenelse{\equal{#1}{Aarno Labs}}{\aarno}{%
		\ifthenelse{\equal{#1}{Galois}}{\galois}{%
			\ifthenelse{\equal{#1}{Intel Corporation}}{\intel}{%
				\ifthenelse{\equal{#1}{University of Washington}}{\wash}{%
					\ifthenelse{\equal{#1}{University of Wisconsin}}{\wisc}{%
						\ifthenelse{\equal{#1}{Yale University}}{\yale}{#1}}}}}}%
}

\newcommand{\RQref}[1]{\hyperref[rq:#1]{\textbf{RQ#1}}}

\date{}
\author{Hamed Okhravi \\
		Brandt Ogden \\
	Noah Luther \\
	Ian McQuoid \\
	Howard Mak \\
	Nathan Burow \\
}
\title{TRACTOR Benchmark for Evaluating C to Rust Translators}

\sponsor{DISTRIBUTION STATEMENT A. Approved for public release. Distribution is unlimited.\\\\
This material is based upon work supported by the Under Secretary of Defense for Research and Engineering under Air Force Contract No. FA8702-15-D-0001 or FA8702-25-D-B002. Any opinions, findings, conclusions or recommendations expressed in this material are those of the author(s) and do not necessarily reflect the views of the Under Secretary of Defense for Research and Engineering.
\\\\
\textcopyright 2026 Massachusetts Institute of Technology.\\\\
Delivered to the U.S. Government with Unlimited Rights, as defined in DFARS Part 252.227-7013 or 7014 (Feb 2014). Notwithstanding any copyright notice, U.S. Government rights in this work are defined by DFARS 252.227-7013 or DFARS 252.227-7014 as detailed above. Use of this work other than as specifically authorized by the U.S. Government may violate any copyrights that exist in this work.}

\begin{document}
\maketitle



%

\tableofcontents

\section{Introduction}
Grace Hopper famously warned that ``the most dangerous phrase in the language is: 'We’ve always done it this way.' '' Nowhere is this observation more relevant than in the construction of critical software.

For decades, many of the world’s most important software systems have been implemented in memory-unsafe languages such as C and C++~\cite{tiobe2026}. These languages underpin operating systems, communication stacks, cryptographic libraries, embedded controllers, and a wide range of safety- and mission-critical applications. Their performance characteristics and low-level control made them compelling historical choices. Yet their weaknesses are equally well understood~\cite{szekeres2013eternal}. Undefined behavior, memory corruption, buffer overflows, and use-after-free errors remain among the most common and most damaging sources of software vulnerabilities~\cite{memsafe}. Despite this persistent risk, vast legacy codebases, performance sensitivities, certification constraints, and ecosystem dependencies have rendered wholesale rewrites largely impractical. As a result, many high-risk systems remain anchored to decades-old implementation decisions.

The urgency of addressing memory safety is not new~\cite{alephone1996smashing}. For more than twenty years, researchers, security practitioners, and industry leaders have repeatedly identified memory-unsafe languages as a systemic root cause of exploitable vulnerabilities~\cite{microsoft2019trends}. Academic studies, breach analyses, and exploit campaigns have consistently traced failures back to memory corruption and undefined behavior~\cite{socprime2026cve20700}. What is unusual about the present moment is not recognition of the problem, but the breadth of agreement about its implications. Technical agencies such as the NSA and CISA~\cite{NSACISAMemorySafe2025}, national policy bodies~\cite{ONCD2024}, and even consumer advocacy organizations~\cite{consumerreports2023memorysafety} have converged on a similar conclusion: incremental mitigation is insufficient, and structural change at the language level is required~\cite{cisa2025securebydesign}.

The software ecosystem has also explored large-scale language modernization before. Efforts such as IBM's Cobol-to-Java modernization work more than two decades ago~\cite{TerekhovVerhoef2000} demonstrated the feasibility of large-scale translation, but they operated in a very different technical and policy environment. Those efforts focused primarily on platform modernization and maintainability rather than on eliminating systemic security vulnerabilities at their root.

Indeed, over the intervening decades, the community invested heavily in attempting to compensate for memory-unsafe languages without abandoning them. Compiler-based mitigations such as stack canaries, address space layout randomization, and control-flow integrity~\cite{cowan1998stackguard, shacham2004effectiveness, burow2017controlflow} were introduced to detect or constrain exploitation. Dynamic defenses such as sandboxing and runtime instrumentation were deployed to limit damage~\cite{lefeuvre2025sokcompartmentalization}. Static analysis tools, verification techniques~\cite{song2019soksanitizing}, hardened allocators~\cite{reitz2024starmalloc}, and secure coding standards such as MISRA and CERT were developed to reduce such vulnerabilities~\cite{misrac, certc}. Each of these advances improved resilience in specific contexts. Yet none eliminated the underlying classes of vulnerabilities. Memory corruption continues to appear across browsers, operating systems, network services, and embedded platforms~\cite{googleprojectzero2014}. In this sense, the shift toward memory-safe languages is not a first attempt at remediation, but the culmination of repeated partial solutions that have not fundamentally resolved the problem~\cite{watson2025time}.

In recent years, however, a qualitatively different path has emerged. Advances in safe programming languages, language tooling, and artificial intelligence have opened the possibility of automated or semi-automated translation of legacy code into safer languages with strong correctness and security guarantees, most notably Rust~\cite{rustbook, c2r1, c2r2, c2r3, c2r4, c2r5, c2r6, c2r7}. Rather than rewriting systems from scratch, these approaches aim to mechanically transform existing implementations while preserving behavior and strengthening safety properties. This line of work draws on static and dynamic program analysis, formal methods, program synthesis, and increasingly large language models. Together, these techniques offer a path to scale beyond what manual refactoring or isolated rewrites can realistically achieve.

Rust~\cite{rustbook2018} occupies a uniquely attractive niche within this modernization landscape. Unlike managed languages that depend on garbage collection, Rust enforces memory safety through compile-time ownership and borrowing rules without imposing runtime GC overhead, preserving deterministic performance characteristics required in systems and real-time domains. It interoperates directly with C and C++ through stable foreign function interfaces, enabling incremental migration rather than all-or-nothing rewrites. It delivers performance comparable to C and C++, in part because it leverages the same LLVM compiler infrastructure and provides explicit control over memory layout and representation. These properties make Rust not merely a safe language for new development, but a technically viable and strategically compelling target for large-scale automated translation.

This broader shift has increasingly been reflected in national policy. In 2024, the White House Office of the National Cyber Director issued Back to the Building Blocks: A Path Toward Secure and Measurable Software, explicitly encouraging the adoption of memory-safe languages and reduction of reliance on inherently memory-unsafe ones~\cite{ONCD2024}. In 2025, the NSA and CISA jointly released Memory Safe Languages: Reducing Vulnerabilities in Modern Software Development, emphasizing the role of language-level safety in improving software resilience and outlining practical pathways for adoption~\cite{NSACISAMemorySafe2025}. These statements build upon years of technical consensus and frame memory safety as a national security priority rather than merely a best practice~\cite{watson2025time}.

Against this backdrop, the DARPA TRACTOR program~\cite{darpa2024tractor} represents a concerted research effort to fundamentally change how legacy, memory-unsafe software is modernized. TRACTOR seeks to develop scalable, automated techniques for translating large C codebases into memory-safe Rust. Although fully automated translation is an aspirational goal, experience suggests that human judgment remains essential. The program therefore frames modernization as a repeatable, measurable, and security-driven process that incorporates human oversight when necessary to ensure correctness and security. Its goals extend beyond functional correctness to include strengthened safety guarantees, competitive performance, and code quality suitable for long-term maintenance and evolution. In doing so, TRACTOR directly addresses the longstanding technical challenge—and growing policy imperative—of reducing systemic dependence on memory-unsafe software.

MIT Lincoln Laboratory serves as the independent test and evaluation (T\&E) organization for the TRACTOR program and is responsible for developing the program's evaluation methodology, benchmarks, metrics, and supporting infrastructure. A central component of this effort is the development and periodic release of standardized benchmark batteries for evaluating C-to-Rust translation tools.

Each battery consists of tens of C test cases curated or synthesized by MIT LL to systematically exercise specific C language features and translation challenges. The batteries are designed to increase in complexity over time, with each release introducing additional features and interactions that translation tools must address. This structure provides a common and progressively challenging benchmark for assessing C-to-Rust translators across functional correctness, safety, idiomaticity, and performance.

The program also includes milestone projects that are substantially larger and more complex than the individual test cases. These projects approximate realistic legacy software components, stressing translation systems in ways smaller examples cannot.

This report describes the TRACTOR benchmark for C-to-Rust translation, developed by MIT Lincoln Laboratory, and covers both the test batteries and milestone projects. It will be updated as new batteries and milestone projects are released. As of September 2026, two batteries (Battery 01 and Battery 02) and three milestone projects (Project 00, Project 01, and Project 02) have been released.

All TRACTOR test batteries and milestone projects are publicly available at:
\url{https://github.com/DARPA-TRACTOR-Program/PUBLIC-Test-Corpus}.

This report also gives an overview of the infrastructure and metrics used to evaluate translators against the benchmark. All associated infrastructure scripts, metrics, and automated evaluation containers are publicly released at: \\ \url{https://github.com/DARPA-TRACTOR-Program/PUBLIC-aws-translate}.

This report does not include evaluation results; it is intended as a reference for C-to-Rust translator developers. Evaluation results for TRACTOR performers are available at:
\url{https://www.ll.mit.edu/tractor}.


\section{Evaluation Metrics and Measurement Approach}
The metrics described here are intended to capture the key dimensions relevant to evaluating a C-to-Rust translation. These dimensions include functional correctness, safety, performance, and idiomaticity, listed in approximate order of importance.

\subsection{Functional Correctness}

A Rust translation is \emph{functionally correct} if it produces the same observable
input/output behavior as the original C program on all test vectors: matching standard
output, standard error, return code, and---for library targets---the expected output
state exposed through the C~ABI.
This is an \emph{empirical}, test-based notion of correctness, not a formal one;
we make no claim of semantic equivalence in the logical or proof-theoretic sense,
and correctness is bounded by the coverage of the test suite.

Undefined behavior (UB) in C complicates this definition.
The C standard permits a compiler to do anything when UB is encountered, so there is
no single ``correct'' output to match.
Nevertheless, for the purposes of this evaluation we treat correctness as
\emph{behavioral fidelity to the compiled C binary}: if the C code, as compiled by
\texttt{clang}/LLVM on the evaluation platform, deterministically produces a specific
output for a given input---even when that output arises from UB---we expect the Rust translation to produce that same
output.
In other words, \emph{C means what LLVM does}: the reference behavior is not the
abstract C standard but the concrete execution produced by the LLVM toolchain on the
test system.
A translation that silently changes the observable output is
considered incorrect for the purposes of this metric; conversely, a translation that preserves the UB-derived output is considered correct.
Test vectors that exercise known UB are currently excluded from automated scoring
(\texttt{has\_ub} flag), and their handling is assessed separately and informally;
improved automated characterization of UB is planned for future evaluation cycles.

To evaluate functional correctness, we curated \emph{test cases}---C programs/projects
collectively referred to as our \emph{test corpus}---in two ways. Our first
method was isolating smaller parts of larger \emph{organic} projects to
evaluate translators on a real-world C code. The second was generating (either
by hand or with the help of a \ac{LLM}) generated \emph{synthetic} projects to
test certain features that could not be easily found or replicated in
real-world programs. Test cases are further divided between \emph{public} and
\emph{hidden}, where public tests cases were released to performers ~6 months
before the evaluation, and hidden test cases were held back and the performers
had no advanced knowledge of.

Another important distinction between test cases is the split between those
which generate executables vs. a shared library (\texttt{.so} file). Test cases
meant to produce a shared library are marked with \texttt{\_lib} at the end of
their name to signify this. The split is important because each lends itself to
a certain style of translation. Executable test cases allow for heavier
optimizations, allowing them to refactor function signatures and the structure
of the project, with the only requirement to maintain the \ac{I/O} behavior
present in the original C executable. A library test is slightly different as
it requires the preservation of symbols and signatures present in the original
C library. This makes optimizations more difficult and requires the use of
\texttt{unsafe}, but allows the Rust library to be a drop-in replacement of the
C library---an important property for a gradual translation from C to Rust.

To accompany each test case we generated (either by hand or with the help of a
\ac{LLM}) \emph{test vectors} that represent the \ac{I/O} behavior that both
the C and Rust projects must adhere to. This necessitated the creation of a
specialized \ac{JSON} schema designed to capture everything involved with a
running executable or library. \autoref{table:vector-schema} outlines each
field in the schema, along with its purpose (is it used for executables,
libraries, or both). \texttt{stdout} and \texttt{stderr} allow for more
permissive matching using regular expressions if specified (e.g., floating
point comparisons). 

\begin{table}[ht]
	\centering
	\caption{Test vector schema}
	\label{table:vector-schema}
	\begin{tabular}{@{}lll@{}}
		\toprule
		\textbf{Field} & \textbf{Description} & \textbf{Purpose} \\ \midrule
		\texttt{argv} & command-line arguments & executable \\[1ex]
		\texttt{stdin} & input provided to program & executable/library \\[1ex]
		\texttt{env} & environment variables passed to program & executable/library \\[1ex]
		\texttt{stdout} & expected output from program & executable/library \\[1ex]
		\texttt{stderr} & expected output to standard error & executable/library \\[1ex]
		\texttt{rc} & expected return code & executable \\[1ex]
		\texttt{lib\_state\_in} & provided input state (see \hyperref[par:cando]{\textbf{Cando}}) & library \\[1ex]
		\texttt{lib\_state\_out} & expected output state (see \hyperref[par:cando]{\textbf{Cando}}) & library \\[1ex]
		\texttt{has\_ub} & description of undefined behavior & executable/library \\[1ex]
		\bottomrule
	\end{tabular}
\end{table}

In addition to the simpler JSON format, we have a second expanded form of a
test vector meant to handle filesystem state equivalence. Our evaluation
infrastructure allows both the new and old formats, and is meant to evolve as
the notion of ''equivalence`` expands as we introduce more complicated
features, particularly networking.

It is a directory in the following form:

\dirtree{%
	.1 <test vector name>/.
	.2 cando\_vector.json.
	.2 setup.
	.2 resources/.
	.2 file\_changes.tar.gz.
}

\texttt{cando\_vector.json} is simply the exact JSON specification defined
above. \texttt{setup} is an arbitrary script that can do any setup that is
required for a test vector to run. It can do anything from creating
files/directories, repositories, or setting up network interfaces.
Additionally, it can set a working directory that the test vector will be run
from. \texttt{resources} is a directory that holds test resources that can be 
accessed from the \texttt{setup} script. \texttt{file\_changes.tar.gz} is 
a tar archive that contains the files that have been created, modified, or deleted
during the run of that test vector. This expanded definition of a test vector
allows us to define equivalence between a C and Rust program in a more robust way,
that is able to evolve as the scope of translations do too.

To facilitate automated function equivalence evaluation we developed two tools,
\texttt{Test Runner}\footnote{Publicly released at
\url{https://github.com/DARPA-TRACTOR-Program/PUBLIC-Test-Corpus/tree/main/tools/test_runner}}
and \texttt{Cando} \footnote{{Publicly released at
\url{https://github.com/DARPA-TRACTOR-Program/PUBLIC-Test-Corpus/tree/main/tools/cando2}}}
. \texttt{Test Runner} facilitates the higher-level orchestration of
\texttt{Cando}, manages filesystem state equivalence, and handles reporting,
while \texttt{cando} evaluates functional equivalence at the test vector level
utilizing the JSON test vector defined above.

\subparagraph{Cando}\label{par:cando} For library tests, \texttt{Cando} uses
the \texttt{lib\_state\_in} and \texttt{lib\_state\_out} fields of the schema
to represent the \emph{state} to pass into, and expect out of the library,
respectively. These are used by the dedicated \emph{harness} (or \emph{runner})
for each library test case that defines how their elements are used. For
example, they could be function arguments, return values, or anything else that
is used by the harness that could define equivalence. The \emph{harness}
specifies the shared library (a \texttt{.so}) file---generated either by C or
Rust---and a \emph{symbol} (function) to call that should exhibit such I/O
behavior. For binary tests, there is a single \emph{harness} that is used
for all binary tests that uses the same JSON specification. The current
implementation skips any test vector marked with \texttt{has\_ub},
though better automatic characterization of UB will be a future improvement.

\subparagraph{Test Runner}\label{par:test-runner} This is a Python project that
is used both to ensure that the test vectors we generate actually match the I/O
behavior of the C test cases we produce (run with Continuous Integration (CI)
on every PR), and runs on Rust projects to check if they conform to the I/O behavior
of the original C (to the extent that the test vectors cover all possible program states).
The following gives a high level overview of what the \texttt{Test Runner} is responsible for:
\begin{itemize}
	\item Discover the C or Rust projects
	\item Build those projects
	\item Build the runners/harnesses needed to run them
	\item Start \texttt{Falco}~\cite{falco}, a runtime security tool that we use to track file changes
	\item Starting and managing test vector containers
	\item Within those containers running the \texttt{setup} script, \texttt{Cando}, and using the file changes from \texttt{Falco} to evaluate filesystem state equivalence
	\item Handling any errors that may occur throughout the run, and report the results to the console and/or to a file as JUnit XML
\end{itemize}

\subsection{Safety}
Safety is a complimentary goal to functionality in that neither is complete
without the other. While it is important for a C-to-Rust translation to be
functionally equivalent to the original C program, it is also important for
that translation to add additional security guarantees that are typical from a
native Rust program. A primary goal for the TRACTOR program is to produce
maximally-safe Rust programs, and simply translating C into entirely
\texttt{unsafe} Rust to evade the borrow checker does not yield any additional
safety benefits. We evaluate Rust translations for safety in two ways, the use
of the \texttt{unsafe} keyword, and \ac{UB}.

\subparagraph{Unsafety} Usefully analyzing and quantifying the use of unsafe
constructs within Rust code is a subtle question. We capture the amount of
unsafety \emph{semantically}, or based on the operations that actually
\emph{require} an unsafe context as opposed to simply being in an
\texttt{unsafe} block. Our semantic analysis\footnote{Publicly released at
	\url{https://github.com/DARPA-TRACTOR-Program/PUBLIC-Test-Corpus/tree/main/tools/rust_eval/unsafe_ops}.}
instruments the Rust compiler to produce a report of these operations that include
calls to unsafe functions, use of inline assemble and mutable statics, raw pointer
dereferences, and more.

Even this instrumentation is unable to perfectly capture unsafety. While we can
capture the number of operations which require unsafety, we cannot determine
whether that unsafety was necessary to implement a particular feature. There
are certain structures which require \texttt{unsafe} contexts, including FFI
boundaries and inline assembly. Library test case translations are required to
be ABI-compatible with the original C, which necessitates the inclusion of
unsafety to expose the required symbols. This is complicated by the fact that
these exposed interfaces can be both implemented directly as \texttt{unsafe}
functions or as wrappers around safe internal functions. Determining the
difference between a ``wrapper'' that simply translates C types into native
Rust types and calling an internal safe function vs. also implementing program
logic, is not something that can be automated but instead requires higher-level
reasoning. In at least one case, we observed a team produce a safe internal
function, and an \texttt{unsafe} wrapper which completely implemented that
function and \emph{did not actually call the internal safe function}. On the
other hand, binary tests have no such requirements to use unsafety to maintain
ABI-compatibility, but there still may remain reasonable or required uses of
\texttt{usafe} that we cannot differentiate between. In this report, we will
report the raw unsafety numbers alongside descriptions of the character of the
unsafety present.

\subparagraph{Undefined Behavior}

Undefined behavior (UB) in C\footnote{See the ISO WG14 report on
	UB~\cite{iso-wg14-ub}.} takes many forms, from classical memory safety errors
to more nuanced issues like the order arguments are evaluated in a function
call. For the purposes of this evaluation, we consider exploitable forms of
undefined behavior, and in particular focus on memory related issues. Removing
vulnerabilities is a key motivation for translating to Rust, so here we assess
how well translators are handling commonly exploited forms of UB.

To assess the handling of UB, we consider two key questions: whether a translator recognizes the presence of UB and, if so, how it handles it. We break this question
down into two parts.  The first is whether to preserve the application's
semantics, or alter them, e.g., by skipping inputs that would result in a
memory corruption, or by adding in manual error handling. The second key
decision is whether to translate to safe code or unsafe code. Safe code
provides runtime checking to detect UB --- though some checks on arithmetic
operations are only present when the code is compiled in debug and not release
mode\footnote{Tooling also exists to catch errors during intermediate stages of
	compilation, for instance Miri~\cite{jung2026miri}.}. Unsafe code does not
typically provide detection of UB.  

When evaluating whether semantics are preserved, the worst option is to
silently change the semantics of a function.  While doing so would ideally be
caught by correctness tests, it is nonetheless dangerous. A better option is
preserving the UB, which, while it decreases the safety gained from translation,
at least has the property of matching behavior. A competing approach is to add
some type of error handling -- either explicitly in the code, likely resulting
in an error message and a crash, or relying on Rust's built in error handling
by translating to safe code. Overall, the best way to handle UB is unclear,
though changing the behavior of the application without developer intervention
seems to be the most dangerous option.

\FloatBarrier

\subsection{Idiomaticity}

Idiomaticity in Rust refers to the extent to which code follows established Rust conventions and leverages the language's intended abstractions, such as ownership and borrowing, pattern matching, and standard library constructs. High idiomaticity is important because it improves readability, maintainability, and safety, and it allows the code to better benefit from Rust's compile-time guarantees and ecosystem tooling.

 While idiomaticity is inherently qualitative and can vary across reviewers, we ground our evaluation rubric in established Rust practices and prior literature. The rubric combines automated measures, including compiler diagnostics and Clippy~\cite{clippy} lints, with structured manual source-code review. This combination is important because neither approach alone provides a sufficient measure of idiomaticity. Manual code-quality assessments can exhibit reviewer inconsistency~\cite{kirk-2025}, while raw linter counts and other static-analysis measures do not necessarily align with human judgments of code quality. Weimer et al.~\cite{weimer-2019}, for example, report weak negative correlations between raw linter warning counts and human rubric scores across four cohorts, and Dehghan et al.~\cite{rustine-2025} similarly caution against using Clippy counts alone to assess idiomaticity.

We release the automated evaluation infrastructure alongside the benchmarks and document the manual evaluation rubric in detail. Together, these resources provide benchmark users with a reproducible framework for assessing the idiomaticity of C-to-Rust translations and for approximating the evaluation methodology used by the TRACTOR Test \& Evaluation team.

\subsubsection{Automated Idiomaticity Evaluation}
Our automated idiomaticity metrics\footnote{Publicly
	released at
	\url{https://github.com/DARPA-TRACTOR-Program/PUBLIC-Test-Corpus/tree/main/tools/rust_eval/static/static_evaluation.py}.}
use both \texttt{rustc} errors/warnings and \texttt{clippy}~\cite{clippy}
lints. \texttt{rustc} errors prevent a Rust program from compiling, which is
already computed by our functional equivalence metric, but the specific reasons for each error are
captured here. Warnings represent potential issues, but do not rise to the level
of a failing compilation. These include unused variables, dead code, and
others, which are important for idiomatic Rust code.

\paragraph{Clippy}
\texttt{clippy} is a dedicated linter for Rust that provides additional best
practices, code style, and idiomatic usage recommendations beyond that of
\texttt{rustc}. There are many different lints available of which we utilize
all in the 5 groups listed below. Each group has examples, though this list is
far from exhaustive.

\begin{itemize}
	\item \textbf{Correctness:} serious issues regarding correctness (either wrong or useless---no false positives)
	\begin{itemize}
		\item \texttt{approx\_constant}: floating point literals close to those already defined (e.g., 3.14)
		\item \texttt{infinite\_iter}: iteration that is guaranteed to be infinite
	\end{itemize}
	\item \textbf{Suspicious:} similar to correctness but possible weird construct is intentional
	\begin{itemize}
		\item \texttt{swap\_ptr\_to\_ref}: calls to \texttt{core::mem::swap()} with parameter derived from pointer
		\item \texttt{empty\_line\_after\_outer\_attr}: empty lines between attribute (e.g., \texttt{\#[allow(dead\_code)]}) and its corresponding function
	\end{itemize}
	\item \textbf{Complexity:} suggestions to simplify constructs (make them shorter and/or readable)
	\begin{itemize}
		\item \texttt{precedence}: identifies operations where precedence is unclear (i.e., readability would be improved with parentheses)
		\item \texttt{unnecessary\_cast}: casts to same type, int/float literals to integer/float types
	\end{itemize}
	\item \textbf{Perf:} hints to increase performance
	\begin{itemize}
		\item \texttt{cmp\_owned}: conversion to owned value only for comparison
		\item \texttt{large\_const\_arrays}: large \texttt{const} arrays that should be defined \texttt{static}
	\end{itemize}
	\item \textbf{Style:} subject style guides for idiomatic code
	\begin{itemize}
		\item \texttt{missing\_safety\_doc}: missing doc comments for \texttt{pub unsafe} functions
		\item \texttt{needless\_return}: use of \texttt{return} at end of block
	\end{itemize}
\end{itemize}

\paragraph{Clippy is not always right}
Clippy's findings can be ignored by an in-source attribute both for a specific
instance via \texttt{\#[allow(...)]} and for many instances (e.g., across an
entire file) via \texttt{\#![allow(...)]}. The commonly accepted procedure for
such overrides is to accompany each attribute with a justification comment
explaining the reason for the violation similar to using safety comments when
using \texttt{unsafe}. 

\Cref{lst:clippy-pi} shows an example of an acceptable use of the
\texttt{\#[allow(clippy::approx\_constant)]} attribute. However, we have also
seen whole project allows that obscure \texttt{clippy} from reporting certain
kinds of issues that would be relevant to the evaluation. Our current automated
idiomaticity metric does not have a way to distinguish between these
situations, and as such the actual number of relevant \texttt{clippy} lints can
be ''gamed`` by overusing \texttt{allow} attributes. Our manual idiomaticity
analysis (described below) takes this into account, and better automatic
characterization of such instances will be improved in future evaluations. 

\begin{lstlisting}[
    caption={Acceptable use of \texttt{\#[allow(clippy::approx\_constant)]} to match C \texttt{M\_PI} macro},
	label={lst:clippy-pi},
    style=ruststyle,
	captionpos=b,
	float=tb
] 
    // GATE-JUSTIFY(clippy::approx_constant): translation requires byte-identity to
    // -O0 C with M_PI resolving to the glibc double macro (3.141592653589793); the
    // literal is intentionally not `std::f32::consts::PI` because the C build
    // evaluates angle math in f64. See q_math.rs module header.
    #[allow(clippy::approx_constant)]
    pub const M_PI: f32 = 3.14159265358979323846_f32;
\end{lstlisting}

\paragraph{Cyclomatic Complexity}
Further, we use \texttt{Lizard}~\cite{lizard} to analyze the Cyclomatic
Complexity (CCN) of both the original C and translator generated Rust. The CCN
is a metric for understanding how many paths there are through each function.
As such, it represents a reasonable proxy for complexity and readability. For
example, a function with no control-flow structures (e.g., \texttt{if}
statements, and \texttt{for} loops) have a CCN of 1. For each additional
control-flow structure added to the function the, CCN increases by 1. We use
the average CCN across all functions in the project under test.

\subsubsection{Manual Idiomaticity Evaluation Introduction}
As with any manual evaluation, idiomaticity assessment is subject to human error, ambiguity, and evaluator bias. To mitigate these effects, we recommend that evaluators first examine the original C code and establish a shared understanding of how a skilled Rust developer would translate it before independently assessing the resulting Rust translation. Evaluators should document the rationale for each score with specific observations from the source code and resolve significant scoring disagreements through discussion. This process is intended to promote consistent application of the rubric while preserving the expert judgment necessary to assess qualities that are difficult to capture through automated metrics alone.

To introduce the notion of a manual evaluation, consider the figures below.
~\autoref{fig:hello-world} shows three different \emph{hello world} programs.
~\autoref{fig:hello-c}, shows the C version that would be given to a
translator. ~\autoref{fig:hello-rust} shows the expected translation into
Rust---what a skilled Rust developer would do. This would get high scores across the board
because it's equally as readable as the original C, uses Rust syntax (i.e.,
the \texttt{println} macro), and uses Rust types (i.e., a string literal) In
contrast, ~\autoref{fig:hello-c2rust} shows the output of C2Rust. This
would score poorly in readability because it turned what could've been
accomplished in 3 lines into 11, doesn't have \texttt{use} statements, and uses
\texttt{unsafe}, all of which obfuscate the intent of the program. Similarly,
it would score poorly in both syntax and types because it directly uses
\texttt{printf} and the C types \texttt{c\_char} and \texttt{c\_int}, respectively.
\FloatBarrier

\begin{figure}
	\centering
	
	\begin{subfigure}[b]{0.4\textwidth}
		\centering
		\begin{lstlisting}[style=cstyle]
#include <stdio.h>
   int main() {
   printf("Hello World\n");
}
		\end{lstlisting}
		\caption{Hello world in C}
		\label{fig:hello-c}
	\end{subfigure}
	\hspace{0.5cm}
	\begin{subfigure}[b]{0.4\textwidth}
		\centering
		\begin{lstlisting}[style=ruststyle, framextopmargin=5.5pt, framexbottommargin=5.5pt]
fn main() {
   println!("Hello World");
}
		\end{lstlisting}
		\caption{Hello world in Rust}
		\label{fig:hello-rust}
	\end{subfigure}
	
	\vspace{0.5cm} 
	\begin{subfigure}[b]{0.9\textwidth}
		\centering
		\begin{lstlisting}[style=ruststyle]
#![allow(dead_code, mutable_transmutes, non_camel_case_types, 
non_snake_case, non_upper_case_globals, unused_assignments, unused_mut)]
extern "C" {
   fn printf(_: *const libc::c_char, _: ...) -> libc::c_int;
}
unsafe fn main_0() -> libc::c_int {
   printf(b"Hello World\n\0" as *const u8 as *const libc::c_char);
   return 0;
}
pub fn main() {
   unsafe { ::std::process::exit(main_0() as i32) }
}
		\end{lstlisting}
		\caption{Hello world from C2Rust}
		\label{fig:hello-c2rust}
	\end{subfigure}
	\caption{Hello world examples of C-to-Rust translations}
	\label{fig:hello-world}
\end{figure}

\subsubsection{Manual Idiomaticity Evaluation Rubric}

We use a five-level Likert scale to provide sufficient differentiation among translations while maintaining a consistent interpretation across idiomaticity dimensions. As shown in Table~\ref{tab:likert-idiomaticity}, levels 1, 3, and 5 serve as ``anchor'' levels corresponding to anti-idiomatic, developing, and idiomatic Rust, respectively, while levels 2 (Novice) and 4 (Proficient) capture translations that fall between these anchors. When a translation partially satisfies the descriptors of two adjacent anchor levels, the corresponding intermediate level should be assigned rather than forcing the assessment to the nearest anchor.

Because idiomaticity requires human judgment, we define a structured rubric across six core dimensions to make explicit \textit{what} we mean by ``idiomatic Rust'' and to promote consistency across evaluators. The five Likert levels intentionally use broad, common descriptors that can be applied uniformly across all six dimensions, from ownership and data flow to project structure and API design. The following sections refine these general descriptors for each dimension by identifying characteristic anti-idioms, expected practices, and signals of highly idiomatic Rust. Together, these dimensions provide a structured framework for assessing translation quality beyond what can be captured through automated metrics alone.

\begin{table}[htbp!]
	\centering
	\caption{Idiomaticity Likert Scale}
	\label{tab:likert-idiomaticity}
	\begin{tabularx}{\textwidth}{|c|l|X|}
		\hline
		Level & Label & Character \\
		\hline
		1 & Anti-idiomatic & Multiple documented Rust anti-idioms in this dimension. Would be out-right rejected in code review. \\
		2 & Novice & Common Rust idioms attempted but misapplied at a structural level. A reviewer would ask for substantial rework rather than touch-ups. \\
		3 & Developing & Correct application of the common Rust idioms. Occasional lapses that a code review would catch. \\
		4 & Proficient & Fluent, consistent Rust with small, but not \textit{meaningful}, lapses. Any deviations are justified in code or docs. \\
		5 & Idiomatic & Could serve as a teaching exemplar. Reader learns the Rust idioms by reading the code. \\
		\hline
	\end{tabularx}
\end{table}

\paragraph{D1 --- Project Structure and Cargo Hygiene}
In dimension 1, we ask the question ``Does the translation look like a normal Rust project a maintainer would write?''.

\begin{enumerate}
	\item \textbf{No hygiene, nothing pinned}
	\begin{itemize}
		\item No \texttt{Cargo.toml}, \texttt{cargo fmt-{}-check} fails, \texttt{clippy::correctness}/\texttt{clippy::suspicious} fire
		\item Monolithic file structure with no delineated boundaries
		\item Unused dependencies present
		\item No tests or trivial tests
		\item Every item \texttt{pub}~\cite{four-horsemen}
	\end{itemize}
	\item \textbf{Ad hoc}
	\begin{itemize}
		\item Builds but layout is ad hoc
		\item Formatting inconsistent
		\item Clippy warnings unaddressed
		\item Tests exist but are all in \texttt{main.rs} or the top of a single file
		\item Some \texttt{todo!} / \texttt{dbg!} in shipped code
	\end{itemize}
	\item \textbf{Standard layout}
	\begin{itemize}
		\item Standard cargo layout (\texttt{src/lib.rs} and/or \texttt{src/main.rs}, \texttt{tests/} for integration, \texttt{\#[cfg(test)] mod tests} for unit)
		\item \texttt{Cargo.toml} has \texttt{name} and \texttt{version}
		\item \texttt{cargo fmt -{}-check} and \texttt{cargo clippy -D warnings} clean at the course lint level
		\item \texttt{README.md} present with build/run instructions
	\end{itemize}
	\item \textbf{Deliberate structure}
	\begin{itemize}
		\item Lib + bin split where both apply (logic in \texttt{lib.rs}, thin \texttt{main.rs})
		\item \texttt{[dev-dependencies]} correctly separated
		\item No unused deps (\texttt{cargo machete} and \texttt{cargo udeps} clean~\cite{fulton-2021})
		\item Module style (\texttt{mod.rs} vs \texttt{foo.rs} + \texttt{foo/}) consistent
		\item \texttt{pub(crate)} / \texttt{pub(super)} narrow visibility
		\item No residual \texttt{dbg!} / \texttt{todo!} / \texttt{unimplemented!} in shipped code
		\item \texttt{Cargo.lock} handled per official guidance (included for bins, not for libs) integration tests use fixtures rather than duplicating setup~\cite{edwards-2008}
	\end{itemize}
	\item \textbf{Template-quality}
	\begin{itemize}
		\item \texttt{[features]} used purposefully with an explicit \texttt{default} and \texttt{dep:} syntax for optional deps
		\item Project layout and tooling could serve as a template for other Rust projects within the domain
	\end{itemize}
\end{enumerate}

\paragraph{D1 --- Project Structure Anti-idioms}
Patterns in this category are signals of a level-1 or level-2 translation
\begin{itemize}
	\item \texttt{todo!()} / \texttt{unimplemented!()} / \texttt{dbg!()} in shipped code~\cite{four-horsemen}
	\item Every item \texttt{pub}~\cite{four-horsemen}
	\item Unused dependencies
\end{itemize}

\paragraph{D1 --- Project Structure Idiomatic Signals}
Patterns in this category are signals of a level-4 or level-5 translation
\begin{itemize}
	\item Modules mirror the domain, not the file listing
	\item \texttt{pub(crate)} and \texttt{pub(super)} used to narrow visibility
	\item Unit tests live next to the code they test
	\item integration tests are their own crate under \texttt{tests/}.
	\item \texttt{[features]} used to gate optional deps, with a documented \texttt{default}
\end{itemize}

\paragraph{D2 --- Ownership and Data Flow}
In dimension 2, we ask the question ``Does the
code work \textit{with} ownership rather than around it?''
\begin{enumerate}
	\item \textbf{Fighting the borrow checker}
	\begin{itemize}
		\item Pervasive \texttt{.clone()} to appease the borrow checker~\cite{robati-shirzad-2024}
		\item \texttt{String} / \texttt{Vec<T>} / \texttt{\&Vec<T>} / \texttt{\&String} parameters where \texttt{\&str} / \texttt{\&[T]} would work
		\item \texttt{Arc<Mutex<..>{}>} or \texttt{RefCell} reached for on first sign of sharing~\cite{qin-2020}
		\item \texttt{let mut} on every binding
		\item Explicit \texttt{'a} on every reference
		\item Occasional \texttt{unsafe} used to escape lifetime issues
	\end{itemize}
	\item \textbf{Cloning less, but still reflexively}
	\begin{itemize}
		\item Some cloning is justified but reflexive clones remain
		\item Some \texttt{\&Vec<T>} / \texttt{\&String} in signatures
		\item Interior-mutability primitives introduced without a stated invariant
		\item Unnecessary \texttt{mut} scattered but not universal
	\end{itemize}
	\item \textbf{Borrows preferred, lifetimes parsimonious}
	\begin{itemize}
		\item Borrows preferred over clones
		\item Parameter types use \texttt{\&str} and \texttt{\&[T]} idiomatically
		\item Interior mutability appears only where justified with a comment
		\item Lifetime annotations parsimonious
		\item Occasional lapse (e.g.\ one \texttt{.clone()} where a borrow would have worked)
	\end{itemize}
	\item \textbf{Ownership expresses intent}
	\begin{itemize}
		\item Ownership expresses intent: move-by-value builders, \texttt{Cow<'\_, str>} where the data flow is mixed borrowed/owned
		\item No defensive cloning
		\item Interior mutability is a documented deliberate choice
		\item Lifetime bounds precise and elided where the compiler allows
	\end{itemize}
	\item \textbf{Ownership restructured to fit the problem}
	\begin{itemize}
		\item Ownership restructured to fit the problem rather than reaching for \texttt{Arc<Mutex<..>{}>} first~\cite{corrode-be-simple}
		\item Shared state minimised architecturally
	\end{itemize}
\end{enumerate}

\paragraph{D2 --- Ownership Anti-idioms}
\begin{itemize}
	\item Defensive \texttt{.clone()} to appease
	\texttt{borrowck}~\cite{four-horsemen,robati-shirzad-2024}
	\begin{itemize}
		\item Note that the anti-idiom is \textit{pervasive / reflexive} cloning, not any clone
	\end{itemize}
	\item \texttt{let mut} on values never mutated~\cite{four-horsemen}
	\item \texttt{Arc<Mutex<..>>} / \texttt{RefCell} as first reach for shared state~\cite{four-horsemen,qin-2020}
	\item \texttt{\&Vec<T>} or \texttt{\&String} in parameter position
	\item Explicit lifetimes where elision applies
\end{itemize}

\paragraph{D2 --- Ownership Idiomatic Signals}
\begin{itemize}
	\item Signatures use \texttt{\&str}, \texttt{\&[T]}, and \texttt{impl AsRef<...>} where appropriate
	\item Move-by-value builders (\texttt{self} not \texttt{\&mut self})~\cite{rust-api-guidelines}
	\item \texttt{Cow<'\_, str>} at API boundaries with mixed ownership
	\item Interior mutability restricted to modules that document the invariant
\end{itemize}

\paragraph{D3 --- Error Handling and Unsafe}
In dimension 3, we ask the question ``Does the
code fail well and treat \texttt{unsafe} with the respect it deserves?''
\begin{enumerate}
	\item \textbf{Panics and undocumented unsafe}
	\begin{itemize}
		\item \texttt{.unwrap()} / \texttt{.expect()} / \texttt{panic!} in normal execution paths~\cite{four-horsemen,three-kinds-of-unwrap}
		\item Magic sentinels (\texttt{-1}, \texttt{usize::MAX}) instead of \texttt{Option} / \texttt{Result}
		\item Errors swallowed via \texttt{.ok()} or \texttt{let \_ = ...}
		\item \texttt{unsafe} present without \texttt{// SAFETY:} comments or beyond clear FFI need~\cite{astrauskas-2020}
		\item Raw pointers used where references would work~\cite{emre-2021}
	\end{itemize}
	\item \textbf{Stringly-typed errors, unexplained unsafe}
	\begin{itemize}
		\item \texttt{Result<\_, String>} in signatures~\cite{rust-api-guidelines}
		\item Some \texttt{?} propagation but mixed with \texttt{.unwrap()} in the same function
		\item Unsafe blocks have brief comments but do not state the invariant being upheld
		\item Interior-unsafe (unsafe hidden in a safe function) without a documented boundary~\cite{qin-2020}
	\end{itemize}
	\item \textbf{Propagation with typed errors, unsafe justified}
	\begin{itemize}
		\item \texttt{?} used throughout for propagation
		\item Typed error enum or \texttt{thiserror} in library code
		\item \texttt{.expect("...")} only where an invariant holds and the message states it (the \texttt{panic!()} \texttt{unreachable!()} senses in Gerat's \emph{Three Kinds of Unwrap} taxonomy~\cite{three-kinds-of-unwrap})
		\item No \texttt{.unwrap()} outside \texttt{\#[cfg(test)]}
		\item \texttt{unsafe} blocks scoped narrowly and preceded by a
		\texttt{// SAFETY:} comment
	\end{itemize}
	\item \textbf{Errors compose, unsafe abstracted safely}
	\begin{itemize}
		\item Errors compose (\texttt{From} impls for conversion)
		\item Diagnostic messages carry enough context to debug from a log line
		\item \texttt{unreachable!()} reserved for truly-unreachable branches with prose justification
		\item \texttt{unsafe} minimised and abstracted behind a safe API with invariants documented in the style of \texttt{std}
		\item \texttt{\# Errors} / \texttt{\# Panics} doc sections cover every fallible / panicking public function
	\end{itemize}
	\item \textbf{Failure modes self-evident, unsafe audit-grade}
	\begin{itemize}
		\item Unsafe either absent or forensically documented such that preconditions, postconditions, and the safe-API boundary stated at the level of an external audit~\cite{astrauskas-2020}
		\item Error types educate the reader
		\item The taxonomy of failure modes is self-evident from the public API
	\end{itemize}
\end{enumerate}

\paragraph{D3 --- Error Handling and Unsafe Anti-idioms}
\begin{itemize}
	\item \texttt{.unwrap()} / \texttt{.expect()} in production paths~\cite{four-horsemen,three-kinds-of-unwrap}
	\item \texttt{Result<\_, String>} in library signatures~\cite{rust-api-guidelines}
	\item Magic sentinels instead of \texttt{Option} / \texttt{Result}~\cite{four-horsemen}
	\item \texttt{unsafe} without \texttt{// SAFETY:}~\cite{astrauskas-2020}
	\item Interior-unsafe with no marked boundary~\cite{qin-2020}
	\item Raw pointers where references work~\cite{emre-2021}
\end{itemize}

\paragraph{D3 --- Error Handling and Unsafe Idiomatic Signals}
\begin{itemize}
	\item \texttt{?} chains with a project-wide error enum
	\item \texttt{// SAFETY:} blocks that reference specific documented invariants
	\item \texttt{\# Errors} and \texttt{\# Panics} sections on every public fallible/panicking fn
	\item Custom error types implementing \texttt{std::error::Error} (\texttt{thiserror} is fine)
\end{itemize}

\textbf{Gate rule:} Any \texttt{clippy::correctness} or
\texttt{clippy::suspicious} finding, or any rapx UB-adjacent finding, gates D3 to \(\leq\) 2 (or 1 for UB) until resolved.

\paragraph{D4 --- Types and Abstractions}
In dimension 4, we ask the question ``Does the
type system carry the invariants, and are the abstractions earned?''
\begin{enumerate}
	\item \textbf{The type system is invisible}
	\begin{itemize}
		\item \texttt{bool} and \texttt{Option<bool>} parameters where a purpose-built enum would read better
		\item Magic numbers
		\item Missing \texttt{Debug} / \texttt{Clone} on public types
		\item Ad-hoc \texttt{new\_from\_x} / \texttt{to\_y} methods duplicating what \texttt{From} / \texttt{Into} would give
		\item Stringly-typed APIs (\texttt{\&str} where an enum or newtype belongs)
		\item C-style typedefs / integer flags where an enum belongs~\cite{four-horsemen}
	\end{itemize}
	\item \textbf{Types attempted but not enforced}
	\begin{itemize}
		\item Some enums replace booleans
		\item Some \texttt{From} impls present but manual conversion methods still dominate
		\item Newtypes attempted but their invariants are not enforced by construction
		\item Generics used defensively (``in case I need it later'')
	\end{itemize}
	\item \textbf{Types carry meaning}
	\begin{itemize}
		\item Enums replace flag booleans
		\item Common derives present where they make sense
		\item \texttt{From} and \texttt{Into} for cheap conversions~\cite{rust-api-guidelines}
		\item Newtypes wrap values that carry useful invariants
		\item Generics used only where actually generic
	\end{itemize}
	\item \textbf{Misuse tends toward a compile error}
	\begin{itemize}
		\item Types encode invariants such that misuse is a compile error (typestate / phantom types where warranted)
		\item \texttt{TryFrom} for fallible conversions
		\item Abstractions justified by $\geq$ 2 uses with no premature generics, no \texttt{dyn Trait} for a single-implementor trait, no builder pattern on a two-field struct~\cite{corrode-be-simple}
		\item \texttt{bitflags} for flag sets
	\end{itemize}
	\item \textbf{Type-driven design is the point}
	\begin{itemize}
		\item The API is essentially incapable of being misused
		\item Generics are elegant and their bounds are minimal~\cite{xu-2021}
		\item A reader learns the domain by reading the types
	\end{itemize}
\end{enumerate}

\paragraph{D4 --- Types and Abstractions Anti-idioms}
\begin{itemize}
	\item Over-parameterised builders (\texttt{FileBuilder<T: AsRef\&str>, 'a>})~\cite{four-horsemen}
	\item \texttt{Box<dyn Iterator>} where \texttt{impl Iterator} fits~\cite{four-horsemen}
	\item Single-implementor traits introduced "just in case"~\cite{four-horsemen}
	\item Stringly-typed APIs~\cite{four-horsemen}
	\item Magic numbers, \texttt{bool} parameters~\cite{rust-api-guidelines}
	\item Unsound generic bounds on unsafe wrappers~\cite{xu-2021}
\end{itemize}

\paragraph{D4 --- Types and Abstractions Idiomatic Signals}
\begin{itemize}
	\item Newtypes at construction time enforce the invariant the type name claims~\cite{rust-api-guidelines}
	\item \texttt{From} / \texttt{Into} (or \texttt{TryFrom} for fallible) rather than named conversion methods
	\item Trait bounds are tight and associated types are purposeful
	\item No \texttt{dyn Trait} unless dispatch actually varies
\end{itemize}

\paragraph{D5 --- Expression and Control Flow}
In dimension 5, we ask the  question ``Does the
code read the way experienced Rust reads?''
\begin{enumerate}
	\item \textbf{C control flow in Rust}
	\begin{itemize}
		\item \texttt{for i in 0..v.len()} indexed loops with \texttt{v[i]} inside~\cite{four-horsemen,birillo-2022}
		\item Imperative accumulation where \texttt{.map().sum()} or \texttt{.collect()} would fit
		\item Nested \texttt{match} / \texttt{if let} pyramids
		\item \texttt{if x \{ true \} else \{ false \}} (clippy \texttt{needless\_bool})
		\item \texttt{.unwrap()} inside \texttt{.map()} / \texttt{.and\_then()} closures
		\item Wildcard \texttt{\_ =>} arms in matches on repo-owned enums
		\item \texttt{while} used where \texttt{for} or an iterator adapter would express intent
	\end{itemize}
	\item \textbf{Idioms appear but don't dominate}
	\begin{itemize}
		\item Some iterator adapters used but for-index loops persist
		\item Matches mostly exhaustive but wildcards silence variants the code should handle
		\item Occasional \texttt{needless\_collect} between adapter stages
		\item Functions large but decomposable~\cite{weimer-2019}
	\end{itemize}
	\item \textbf{Idiomatic expression is the default}
	\begin{itemize}
		\item Iterator adapters used where they simplify
		\item \texttt{?} chains where they express intent
		\item Pattern matching leveraged well (\texttt{if let}, \texttt{let ... else}, exhaustive \texttt{match})
		\item Wildcards used only where the enum is open
		\item Cognitive complexity in the neighbourhood of Ardito's \textasciitilde{}0.7/method idiomatic baseline~\cite{ardito-2021}
	\end{itemize}
	\item \textbf{Construct choice is deliberate}
	\begin{itemize}
		\item Appropriate blend of plain \texttt{for} loops when it reads better than a functional chain~\cite{corrode-be-simple}
		\item No needless \texttt{.collect()} between iterator stages
		\item Destructuring in bindings and function arguments
		\item \texttt{matches!} / \texttt{let ... else} used exactly where they improve clarity, not because they exist
		\item Macros used only for meta uses
	\end{itemize}
	\item \textbf{Every construct legible as a choice}
	\begin{itemize}
		\item Reading the code is faster than reading its comments
		\item Complex operations pipeline cleanly
		\item The choice between imperative and functional shape is deliberate every time
	\end{itemize}
\end{enumerate}

\paragraph{D5 --- Expression and Control Flow Anti-idioms}
\begin{itemize}
	\item Indexed loops~\cite{four-horsemen,birillo-2022}
	\item \texttt{needless\_bool}, \texttt{needless\_collect}, \texttt{redundant\_pattern\_matching} clippy warnings
	\item \texttt{.unwrap()} in closures
	\item Catch-all \texttt{\_ =>} arms in matches on enums we declared
	\item Functions over ~30 statements without decomposition~\cite{weimer-2019}
\end{itemize}

\paragraph{D5 --- Expression and Control Flow Idiomatic Signals}
\begin{itemize}
	\item \texttt{.iter().map(...).filter(...).collect()} chains where they compose
	\item \texttt{for x in ...} when a chain would obscure the shape
	\item \texttt{let Some(x) = ... else { return ... };} where a match would be better suited
	\item Cognitive complexity \(\leq\) 1.0 per method~\cite{ardito-2021}
\end{itemize}

\paragraph{D6 --- API, Naming, and Documentation}
In dimension 6, we ask the question ``Would someone else be able to use this crate without reading its source?''

\begin{enumerate}
	\item \textbf{Opaque without the source}
	\begin{itemize}
		\item \texttt{get\_foo()}-style getters~\cite{rust-api-guidelines}
		\item \texttt{///} docs absent on \texttt{pub} items
		\item Examples that use \texttt{.unwrap()} or that don't compile
		\item Missing \texttt{\# Errors} / \texttt{\# Panics} sections
		\item Every item is \texttt{pub}~\cite{four-horsemen}
		\item \texttt{Debug} missing on public types
	\end{itemize}
	\item \textbf{Documented in places}
	\begin{itemize}
		\item Some public items documented but examples don't run
		\item Naming mostly follows RFC 430 but some abbreviations
		\item \texttt{\# Errors} sections missing on half of fallible functions
		\item Ad-hoc conversion methods with unclear semantics
	\end{itemize}
	\item \textbf{Conventional and navigable}
	\begin{itemize}
		\item Collections expose \texttt{iter} / \texttt{iter\_mut} / \texttt{into\_iter}
		\item \texttt{as\_} / \texttt{to\_} / \texttt{into\_} used per their cost conventions
		\item \texttt{///} docs on public items
		\item \texttt{\# Errors} / \texttt{\# Panics} where applicable (\texttt{\# Panics} when a panic is reachable under any input, instead of when a panic-macro token appears in the body
		\item \texttt{panic!()} / \texttt{unreachable!()} in a branch the surrounding control flow proves unreachable does not require a \texttt{\# Panics} section)
		\item Examples use \texttt{?}
		\item \texttt{pub(crate)} used to narrow scope
	\end{itemize}
	\item \textbf{Guided use, runnable docs}
	\begin{itemize}
		\item Constructors are \texttt{new}
		\item \texttt{From} / \texttt{AsRef} over inherent conversion methods
		\item \texttt{\#[must\_use]} where a return is easy to drop by accident
		\item Rustdoc examples runnable and present on the fallible/panicking public surface (\texttt{cargo test -{}-doc} passes and covers more than a single trivial function)
		\item \texttt{Debug} on public types
		\item Struct fields private with accessors where invariants exist
	\end{itemize}
	\item \textbf{The API teaches its own use}
	\begin{itemize}
		\item The public API teaches its own use such that the docs are the mental model
		\item Every export is intentional and its cost is stated
		\item The crate could be released to crates.io without significant editing
	\end{itemize}
\end{enumerate}

\paragraph{D6 --- API, Naming, and Documentation Anti-idioms}
\begin{itemize}
	\item \texttt{///} docs absent on \texttt{pub} items.
	\item \texttt{\# Errors} / \texttt{\# Panics} missing on fallible / panicking public functions
	\item \texttt{get\_foo()} getters (Rust uses \texttt{foo()} and \texttt{foo\_mut()})~\cite{rust-api-guidelines}
	\item Ad-hoc \texttt{new\_from\_x} / \texttt{to\_y} methods where \texttt{From} / \texttt{Into} / \texttt{AsRef} fit
	\item Everything \texttt{pub}~\cite{four-horsemen}
	\item Public types without \texttt{Debug}
\end{itemize}

\paragraph{D6 --- API, Naming, and Documentation Idiomatic Signals}
\begin{itemize}
	\item Doctest examples that use \texttt{?} and are runnable
	\item \texttt{\#[must\_use]} on returns that carry meaning (e.g., \texttt{Result}, builders)
	\item Public structs with private fields and constructor invariants
	\item Module-level \texttt{//!} doc that states the invariant the module protects
\end{itemize}

\subsection{Performance}
\subparagraph{Translation Throughput} 

Translation throughput captures the rate at which a translator can process C code and is an important consideration for practical adoption. We measure throughput in thousands of lines of C code translated per hour (KLOC/hr). Throughput is measured at the granularity of the complete AWS translation job rather than the translator's internal processing time. Consequently, the measurement includes standardized overhead associated with environment setup and teardown and data transfer to and from S3. Because execution time in shared cloud environments can vary due to factors outside the translator's control, throughput measurements should be interpreted with this inherent variability in mind.

Runtime and memory overhead measurements are implemented in \hyperref[par:cando]{\textbf{Cando}} and \hyperref[par:test-runner]{\textbf{Test Runner}}. We additionally provide a higher-level benchmarking script designed to reduce measurement bias and variability. The script collects the relevant (translator, test case, test vector) tuples, randomizes their execution order, executes a randomly selected subset as warm-up runs, and then performs and reports the benchmark measurements. This methodology is intended to reduce systematic effects from execution order, cache state, and other sources of measurement noise, thereby supporting more reproducible comparisons across translators.

\subparagraph{Runtime Overhead} 
Rust adopts the philosophy of zero-cost abstractions, or useful abstractions
that negligibly impact runtime. Successful translations into Rust should
exploit this principle to limit the increase---or even decrease---the time
it takes to execute the same workload in C. We evaluate this
using \texttt{perf-stat}~\cite{perfstat}, a common tool for gathering
performance statistics for C and Rust programs.
This captures several key statistics including, task clock, user, and total runtime, 
instructions executed, CPU cycles, cach references/misses, among others
that are relevant to our evaluation. 

\subparagraph{Memory Overhead} 
Similarly---though not as important on modern hardware where memory is
plentiful---is the goal of a Rust program's memory footprint to be comparable
to the original C. To support this we use another common program,
\texttt{valgrind}~\cite{valgrind}. Its tool \texttt{massif} gives a snapshot of
the number of bytes allocated to the heap, stack, and used as metadata for the
memory allocator, throughout the execution of a program. An important caveat to
note is our use of \texttt{valgrind} can not always capture memory usage
information. Particularly, if a test vectors runs and finishes \emph{very}
quickly, the IPC mechanisms we need to setup will sometimes not initialize fast
enough to send information back. However, this has not been observed very
often, so is not a major limitation affecting our data collection.

\section{Benchmark Batteries and Milestone Projects }

\subsection{Battery 01 Benchmark}

The first test battery, B01, was designed to contain almost exclusively C programs which can be represented easily within the Rust type system.
In particular, it excludes programs which store memory on the heap, excludes programs which exhibit parallel execution, and aims to exclude
programs which exhibit memory access patterns that are difficult to represent in Rust's affine type system (e.g., interior mutability).

The explicit set of features included in B01 are: 
\begin{itemize}
	\item Pointers to data on the stack or static  
	\begin{itemize}
		\item No \texttt{malloc} or \texttt{void*}
		\item Static initialization/constants 
		\item Pointer arithmetic only: 
		\begin{itemize}
			\item Into arrays
			\item Using \texttt{argc} to index/parse \texttt{argv}
		\end{itemize}  
		\item Pass-by-reference for values/structs not involving pointer arithmetic 
		\item Scoped arrays 
	\end{itemize}
	\item Error handling 
	\begin{itemize} 
		\item \texttt{errno.h}
		\item Functions that return Boolean or other values for success/failure, and put the result of computation into a referenced value
		\begin{itemize} 
			\item Many \texttt{libc} functions do this, for instance 
			\item Possibly including different references pointing to members of the same object (difficult for Rust ownership) 
		\end{itemize}
	\end{itemize}
	\item Global state 
	\begin{itemize}
		\item Compile-time constants 
		\item Static variables (mutating during execution) 
	\end{itemize}
	\item Simple data structs without recursion and with size known at compile time
	\item Code using \texttt{libc}
	\begin{itemize} 
		\item Rust translations will require Foreign Function Interfaces (FFI) calling C unless they swap in a Rust library or attempt to translate referenced functions
		\item Functions from \texttt{libc} should (mostly) have Rust analogues that map naturally 
	\end{itemize}
	\item Library functions 
	\begin{itemize}
		\item C interface to test with (as a drop-in replacement) 
		\item This will require unsafe code to permit FFI from C 
	\end{itemize}
\end{itemize}

\subsection{Milestone Projects 00 and 01 Benchmarks} \label{sec:milestones}
To complement the B01 test battery, we include two milestone projects drawn from the open-source community. These projects provide larger, more realistic translation targets that exercise the C features represented in B01 while introducing the additional challenges associated with translating complete software projects. The two projects included in this benchmark are:

\begin{enumerate}
	\item Milestone Project~0 (P00): A repository consisting of the Perlin noise library~\cite{perlin}. P00 is a compact but computation-heavy codebase that exercises floating-point arithmetic, array indexing, and performance-sensitive inner loops making it representative of numerical and signal-processing workloads commonly found in legacy C code.
	\item Milestone Project~1 (P01): A repository focusing on the SPHINCS\texttt{+} cryptographic library~\cite{sphincsplus}, which is substantially larger and more complex than P00, and stresses translation tools with extensive bit-level manipulation, strict correctness requirements, and sensitivity to even minor semantic deviations. P01 additionally tests translation tools' management of build systems producing multiple libraries and supporting multiple configurations.
\end{enumerate}
Together, these milestones complement the test battery by introducing more realistic and domain-specific translation challenges.

Both projects were modified versions of open-source libraries with the original Perlin Noise library authored by Sean Barrett~\cite{perlin} and the original SPHINCS\texttt{+} implementation authored by Eyal Ronen~\cite{sphincsplusc} leveraging the NIST Post-Quantum Cryptography reference implementation~\cite{splincsplusref} along with an implementation of the BLAKE hash function~\cite{aumasson2008sha3} taken from the SUPERCOP benchmark framework~\cite{supercop}. The two projects were modified to only exercise a similar subset of C features to those in B01; however P01 underwent a significant transformation to test the translation tools' handling of C project structure and configurability.

\subsection{Battery 02 Benchmark}  \label{sec:batch-b02}
The second test battery, B02, was designed to admit new classes of C program which the previous battery excluded as resistant to safe Rust representation. B01 largely confined itself to programs expressible within Rust's affine type system while B02 relaxes those restrictions one ring outward by introducing heap-allocated state, shared mutable data, indirect dispatch, and manual resource-cleanup control flow. These constructs are precisely those a translation tool must map onto Rust's ownership, borrowing, and trait machinery to achieve a fully safe translation as opposed to simply producing a direct syntactic analogue. This being said, B02 remains a bounded battery and programs were heavily edited to avoid C features that are out-of-bounds for the battery such as inline assembly, cyclic ownership patterns, and concurrency.

The explicit set of features introduced in B02 are:
\begin{itemize}
	\item Heap allocation and ownership
	\begin{itemize}
		\item Dynamic allocation via \texttt{malloc} / \texttt{calloc} / and \texttt{realloc} with release through \texttt{free}
		\item Transfer of ownership across function boundaries (``allocate-here, free-there'' patterns)
		\item \texttt{void*} payloads and buffers
	\end{itemize}
	\item Mutable aliasing and interior mutability
	\begin{itemize}
		\item Multiple live, mutable paths reaching the same object (difficult for Rust's borrow checker)
		\item Shared mutable state reached indirectly, including through function pointers
	\end{itemize}
	\item Raw buffer manipulation with a focus on strings and the standard library
	\begin{itemize}
		\item Pointer-and-length buffers with possible overlap (\texttt{memcpy}, \texttt{memmove}, and the \texttt{mem*}/\texttt{str*} family)
		\item In-place editing, appending, and search/replace over byte and character buffers
	\end{itemize}
	\item Indirect dispatch
	\begin{itemize}
		\item Function pointers stored as data: operation tables, callback fields, and typed vtable-style backends
		\item Function-pointer values punned through \texttt{void*}
	\end{itemize}
	\item Simple generic containers
	\begin{itemize}
		\item Type-erased hash maps and growable arrays, instantiated through macros (which may be expressed with generics or trait objects in Rust)
	\end{itemize}
	\item Error handling via cleanup control flow
	\begin{itemize}
		\item \texttt{goto}-based cleanup ladders (which may translated into \texttt{Drop} and the \texttt{?} operator in Rust)
	\end{itemize}
	\item Type punning
	\begin{itemize}
		\item Reinterpretation of storage through \texttt{union} members and packed bit-fields
		\item Reinterpreting casts between object representations, \textit{e.g.} serializing structured data to and from byte buffers
	\end{itemize}
	\item Data structures with structural indirection
	\item Program structure
	\begin{itemize}
		\item Cross-translation-unit linkage: internal-linkage (\texttt{static}) symbols, \texttt{extern} redeclaration, \texttt{static inline}, and name reuse across units
		\item Deep preprocessor composition (token pasting, stringization, multi-level, and recursive macro expansion)
	\end{itemize}
	\item File and filesystem interaction
	\begin{itemize}
		\item File I/O through descriptors (open/read/write/close), directory traversal, and path manipulation
	\end{itemize}
	\item Environment-driven state
	\begin{itemize}
		\item Configuration read from the environment (\texttt{getenv})
	\end{itemize}
\end{itemize}

The organic half of B02 is drawn from small third-party codebases rather than from single-purpose fragments: a JSON parser exhibiting the \texttt{goto}-cleanup idiom throughout, two-dimensional collision-detection and ray-casting primitives mirroring compact single-header game libraries stretchy-buffer and hash-map macros, and image-decoding, inflate, and Base64/UTF-8 codecs. A small number of entries are drawn from the Underhanded C Contest~\cite{UnderhandedC} (years 2009 and 2015): programs whose intended behavior is deliberately hard to read from the source, included to stress translation \emph{fidelity} rather than any single language feature. The synthetic half focuses on explicit features or interfaces of multiple features as in the previous battery.

\subsection{Milestone Project 02 Benchmark} \label{sec:project-p02}
As with Milestone Projects P00 (Perlin Noise) and P01 (SPHINCS+), a single open-source project was chosen to combine the battery's isolated features under one build. Milestone Project~2 (P02) is a heavily modified port of the \texttt{libgit2} library~\cite{libgit2}, the linkable implementation of Git. P02 is substantially larger than the earlier milestones at the order of 100-thousand lines of code and was chosen to not only exercise the full B02 feature set at once, but was also chosen as an intermediate point for principled ownership analysis as libgit2 relies on acyclic and reference-based data structures after excision of the internal memory pool. In this sense, P02 is a capstone that escalates the same battery to test the scaling complexity of translation tools and as a core method of differentiating translation based on the Rust idioms or structures the translator chooses.

Together, B02 and its milestone move the evaluation from feature coverage toward translation soundness. The constructs the battery introduces (including mutable aliasing, buffer overlap, lifetime-bearing cleanup paths, indirect dispatch, and deliberately-subtle semantics) are exactly those whose correct translation is hardest to establish by inspection of the source alone as they enable properties that no single site makes locally visible. B02 and P02, therefore, make up the initial evaluation in which a tool's ability to preserve behavior, and not merely parse and re-emit constructs, is placed under test.

\subsection{Evaluation Infrastructure}

The test and evaluation infrastructure\footnote{Publicly released at \url{https://github.com/DARPA-TRACTOR-Program/PUBLIC-aws-translate}.} provides a standardized and reproducible environment for executing C-to-Rust translators against the benchmark. The infrastructure is instantiated using the AWS Cloud Development Kit (CDK) in Python and deploys translators as AWS Batch jobs on Fargate. Each translator is packaged as a Docker container and executed within a common Virtual Private Cloud (VPC), with external connectivity available for translators that rely on LLM API endpoints.

Translator containers are provided read access to an S3 bucket containing the C benchmark inputs and write access to a separate S3 bucket for their generated Rust translations. Each input archive contains the C source code and the materials necessary to build it, but does not include the benchmark test vectors (i.e., sample inputs and expected outputs). Each test case provides either a top-level \texttt{CMakeLists.txt} or \texttt{CMakePresets.json} specifying how the original C program should be built. For translators that rely on external LLM services, API credentials can be securely provided through AWS Secrets Manager. Together, this infrastructure enables researchers to execute and evaluate translators under a common environment while reproducing the benchmark methodology independently.

\section{Conclusion}

The transition from legacy C software to memory-safe Rust presents both a significant opportunity and a difficult technical challenge. Successful translation requires substantially more than producing Rust code that compiles or reproduces the behavior of a C program. A useful translation must simultaneously preserve functional behavior, reduce reliance on unsafe constructs, maintain acceptable performance, and produce idiomatic Rust that can be understood, maintained, and evolved by developers. The TRACTOR benchmark is designed to evaluate these dimensions together rather than treating translation as a purely syntactic transformation. 

To support this goal, the benchmark combines progressively challenging test batteries with larger milestone projects. Battery 01 establishes a foundation around C constructs that map relatively naturally into Rust's type and ownership systems, while Battery 02 introduces substantially harder challenges such as dynamic allocation, mutable aliasing, indirect dispatch, type erasure, cleanup control flow, and complex program structure. The milestone projects complement these targeted tests by evaluating whether translation techniques scale to realistic software, culminating in P02, a substantially larger codebase that stresses both translation scale and the ability to reason about ownership and behavior across program boundaries.   

Equally important, the benchmark defines a reproducible methodology for determining what constitutes a successful translation. Functional correctness is evaluated through observable behavioral equivalence; safety considers both semantic uses of `unsafe` and the treatment of undefined behavior; idiomaticity combines automated analysis with a structured manual rubric; and performance captures translation throughput as well as runtime and memory overhead. Together, these metrics recognize that no single measure adequately captures translation quality.    

The benchmark and its associated evaluation infrastructure are intended to evolve alongside C-to-Rust translation technology. As increasingly difficult batteries and milestone projects are introduced, the evaluation can expand to cover additional language features, system interactions, and notions of behavioral equivalence. By publicly releasing the benchmarks, metrics, and supporting infrastructure, the TRACTOR program provides a common basis for measuring progress and comparing approaches.  Ultimately, the objective is not simply to demonstrate that C can be translated into Rust, but to establish whether such translation can be performed correctly, safely, efficiently, and at the scale necessary to make memory-safe modernization of legacy software practical.

\bibliographystyle{llbib}
\bibliography{references}

\end{document}